\documentclass[12pt]{article}
\usepackage{latexsym}
\usepackage{amssymb}
\usepackage{graphicx}

\def\be{\begin{equation}}
\def\ee{\end{equation}}
\def\bear{\begin{eqnarray}}
\def\eear{\end{eqnarray}}
\def\nn{\nonumber}

\newcommand\bra[1]{{\langle {#1}|}}
\newcommand\ket[1]{{|{#1}\rangle}}

\def\a{\alpha}
\def\b{\beta}
\def\g{\gamma}
\def\d{\delta}

\def\s{\sigma}
\def\th{\theta}

 \def\o{{\rm ord}}
 
 \def\L{{\Lambda}}

 \def\IZ{\relax\ifmmode\mathchoice
 {\hbox{\cmss Z\kern-.4em Z}}{\hbox{\cmss Z\kern-.4em Z}}
 {\lower.9pt\hbox{\cmsss Z\kern-.4em Z}}
 {\lower1.2pt\hbox{\cmsss Z\kern-.4em Z}}\else{\cmss Z\kern-.4em Z}\fi}
 \def\IB{\relax{\rm I\kern-.18em B}}
 \def\IC{{\relax\hbox{$\inbar\kern-.3em{\rm C}$}}}
 \def\Ic{{\relax\hbox{$\inbar\kern-.22em{\rm c}$}}}
 \def\ID{\relax{\rm I\kern-.18em D}}
 \def\IE{\relax{\rm I\kern-.18em E}}
 \def\IF{\relax{\rm I\kern-.18em F}}
 \def\IG{\relax\hbox{$\inbar\kern-.3em{\rm G}$}}
 \def\IGa{\relax\hbox{${\rm I}\kern-.18em\Gamma$}}
 \def\IH{\relax{\rm I\kern-.18em H}}
 \def\II{\relax{\rm I\kern-.18em I}}
 \def\IK{\relax{\rm I\kern-.18em K}}
 \def\IP{\relax{\rm I\kern-.18em P}}

\def\Tr{{\rm Tr}}
 \font\cmss=cmss10 \font\cmsss=cmss10 at 7pt
 \def\IR{\relax{\rm I\kern-.18em R}}

\def\dd{\mbox{d}}

\def\o{\omega}
\def\bra{\langle}
\def\ket{\rangle}
\def\a{\alpha}
\def\b{\beta}
\def\d{\delta}
\def\D{\Delta}

\def\g{\gamma}
\def\G{\Gamma}
\def\e{\epsilon}
\def\ve{\varepsilon}

\def\f{\phi}
\def\F{\Phi}
\def\vf{\varphi}
\def\k{\kappa}
\def\l{\lambda}
\def\L{\Lambda}
\def\m{\mu}
\def\n{\nu}
\def\s{\sigma}

\def\o{\omega}

\def\th{\theta}

\def\pa{\partial}

\newcommand{\ti}[1]{\tilde{#1}}

\newcommand{\sm}[1]{\mbox{\scriptsize #1}}

\renewcommand{\@}[1]{\sqrt{#1}}
\renewcommand{\le}[1]{\label{#1}\end{eqnarray}}
\newcommand{\bea}{\begin{eqnarray}}
\newcommand{\eea}{\end{eqnarray}}

\newcommand{\eq}[1]{(\ref{#1})}
\def\nn{\nonumber\\}

\def\ffract#1#2{\raise .35 em\hbox{$\scriptstyle#1$}\kern-.25em/
\kern-.2em\lower .22 em \hbox{$\scriptstyle#2$}}

\def\na{\nabla}
\def\half{{1\over2}\,}
\newdimen\tableauside\tableauside=1.0ex
\newdimen\tableaurule\tableaurule=0.4pt
\newdimen\tableaustep
\def\phantomhrule#1{\hbox{\vbox to0pt{\hrule height\tableaurule width#1\vss}}}
\def\phantomvrule#1{\vbox{\hbox to0pt{\vrule width\tableaurule height#1\hss}}}
\def\sqr{\vbox{%
  \phantomhrule\tableaustep
  \hbox{\phantomvrule\tableaustep\kern\tableaustep\phantomvrule\tableaustep}%
  \hbox{\vbox{\phantomhrule\tableauside}\kern-\tableaurule}}}
\def\squares#1{\hbox{\count0=#1\noindent\loop\sqr
  \advance\count0 by-1 \ifnum\count0>0\repeat}}
\def\tableau#1{\vcenter{\offinterlineskip
  \tableaustep=\tableauside\advance\tableaustep by-\tableaurule
  \kern\normallineskip\hbox
    {\kern\normallineskip\vbox
      {\gettableau#1 0 }%
     \kern\normallineskip\kern\tableaurule}%
  \kern\normallineskip\kern\tableaurule}}
\def\gettableau#1 {\ifnum#1=0\let\next=\null\else
  \squares{#1}\let\next=\gettableau\fi\next}

\begin{document}

\begin{flushright}
ITP-UU-08/49\\
SPIN-08/39\\
\end{flushright}
\vskip0.5truecm

\centerline{{\Large \bf Dual Gravitons in AdS$_4$/CFT$_3$ and the}} 
\vskip.2truecm
\centerline{{\Large \bf Holographic Cotton Tensor}}


\vskip 2truecm

\begin{center}
{\large Sebastian de Haro}\\
\vskip .4truecm
\vskip 4truemm {\it Institute for Theoretical Physics and Spinoza Institute\\
Utrecht University\\
Leuvenlaan 4, 3508 TD Utrecht, The Netherlands}\\
\tt{sebastian.deharo@gmail.com}\\

\end{center}
\vskip 1.5truecm

\begin{center}

\textbf{\large \bf Abstract}

\end{center}

We argue that gravity theories in AdS$_4$ are holographically dual to either of two three-dimensional CFT's: the 
usual Dirichlet CFT$_1$ where the fixed graviton acts as a source for the stress-energy tensor, and a dual 
CFT$_2$ with a fixed dual graviton which 
acts as a source for a dual stress-energy tensor. The dual stress-energy tensor is shown to be the Cotton tensor 
of the Dirichlet CFT. The two CFT's are related by a Legendre transformation generated by a gravitational 
Chern-Simons coupling. This duality is a gravitational version of electric-magnetic duality
valid at any radius $r$, where the renormalized stress-energy tensor is the electric field and the Cotton tensor is
the magnetic field. Generic Robin boundary conditions lead to CFT's coupled to Cotton gravity or topologically 
massive gravity. Interaction terms with CFT$_1$ lead to a non-zero vev of the stress-energy tensor in CFT$_2$
coupled to gravity even after the source is removed. We point out that the dual graviton also exists
beyond the linearized approximation, and spell out some of the details of the non-linear construction.

\newpage

\tableofcontents

\newpage

\section{Introduction}

In AdS/CFT, the boundary conditions on the two independent modes of the graviton with, respectively, slow and fast 
fall-off at infinity encode the (representative of the conformal class of) boundary metrics and the one-point function 
of the stress-energy tensor of
the CFT \cite{Witten,dHSS}. For four-dimensional pure gravity, this is summarized in the holographic relation:
\be\label{stressT}
\bra T_{ij}(x)\ket={3\ell^2\over16\pi G_N}\,g_{(3)ij}(x)~,
\ee
that is, the third coefficient in the boundary expansion of the graviton $g_{ij}(r,x)$ is the renormalized 
holographic stress-energy tensor (see notation in Appendix \ref{holren}). The stress-energy tensor should be 
seen here as a functional computed as the CFT response to the background boundary metric $g_{(0)}$.

This result is obtained from the identification of the on-shell bulk action with the generating functional of
connected correlation functions in the CFT, leading to:
\be\label{set}
\d S[g_{(0)}]=({\mbox{eom}})+\half\int\dd^3x\,\sqrt{g_{(0)}}\,\bra T_{ij}(x)\ket\,\d g^{ij}_{(0)}(x)~.
\ee
The bulk action, including boundary terms and counterterms, is given in Appendix \ref{holren}. With 
Dirichlet boundary conditions, this gives the definition of the renormalized stress-energy tensor after sending
the regulator to zero.

In the Neumann variational problem, the stress-energy tensor is fixed, in particular equation \eq{set} sets it 
to zero. Mixed boundary conditions are
obtained, in the case of scalar fields, by adding boundary terms that correspond to multiple-trace deformations 
of the CFT. For 
gravity, Neumann and mixed boundary conditions were not studied in the past, except for brane-world scenarios 
\cite{dHSS2}, for at least two reasons:\\
\\
1) The Neumann and mixed variational problems do not always lead to a consistent bulk quantization scheme. This is 
due to the fact that generically the mode with fast fall-off is normalizable whereas the mode with 
slow fall-off is non-normalizable. Therefore, the latter must be fixed at the boundary, as in a Dirichlet boundary 
problem. Only in special cases, such as scalar fields in the range of masses $-{d^2\over4}<m^2<{d^2\over4}+1$, or if 
the theory has a cutoff, are the Neumann and mixed boundary conditions admissible.\\
\\
2) A Neumann variational problem would naively identify the graviton as the dual operator, which does not make 
sense in a CFT$_3$. Indeed, if the quantity that is being held fixed is the canonical momentum --the renormalized
holographic stress-energy tensor-- rather than the 
asymptotic value of the metric, the former will act as a source for the boundary graviton since they 
are coupled in the on-shell bulk action. However, a spin two operator of dimension zero is below the unitarity bound. If the 
CFT is coupled to gravity and the graviton is integrated over, the non-zero expectation value of the graviton would not 
make sense at all, as a gravity theory does not have local operators. At best, it might make sense in some
perturbative sense, where the CFT is expanded about a suitable background.

The resolution of the first problem has been known for some time. Ishibashi
and Wald showed four years ago \cite{IW} that in four dimensions, both the Dirichlet and the Neumann boundary
conditions lead to a well-posed initial value problem by showing that there is a suitable self-adjoint extension
of a certain differential operator which plays the role of the Hamiltonian. This self-adjoint extension determines
the asymptotic boundary conditions, so the freedom in choosing self-adjoint extensions of this operator 
corresponds to the freedom in choosing suitable boundary conditions at infinity. Their analysis showed that in
four dimensions, the self-adjoing extensions are such that Dirichlet, Neumann boundary conditions as well as 
Robin boundary conditions are possible. In the latter case a linear combination of the leading and the subleading 
modes of the graviton is held fixed. 

In this paper we propose a solution to the second problem above, giving the correct holographic interpretation of the 
Neumann and mixed problems. The key question we will address is: what are the observables of the dual theory? Here, it is useful to first
recall the scalar field case in the special range of masses mentioned above, where duality interchanges sources in CFT$_1$ with one-point functions in CFT$_2$, 
and viceversa. Concretely, for Dirichlet quantization the mode $\f_+$ with fast fall-off corresponds to the dual operator $\bra{\cal O}_{(\D_+)}\ket$ of dimension $\D_+$, 
whereas the slow mode is a fixed source $\f_-=J$. In the second quantization scheme, these roles are interchanged
and $\f_-$ corresponds to the one-point function of an operator of dimension $\D_-$ whereas $\f_+$ is held fixed.
Deforming the CFT by a multiple-trace operator, one can flow towards the Dirichlet CFT \cite{multitrace,multitrace1}.

For gravity, the picture is more involved. Generically, the energy-momentum tensor does not depend on the
boundary graviton. However, its expectation value does depend on it. This can be seen from the bulk via the
regularity condition. Now in three dimensions one may classically associate to a given stress-energy tensor a 
graviton, constructed through the Cotton tensor.  Suppose we have CFT$_1$ with fixed graviton $h_{ij}$ and stress-energy
tensor $\bra T_{ij}\ket$. In three dimensions, given $h_{ij}$, we can map it to a conserved, 
traceless object of dimension 3, via the three-dimensional Cotton tensor
(see Appendix \ref{curvtensors}):
\be\label{TC}
\bra \ti T_{ij}\ket=C_{ij}[h]~.
\ee
On the other hand, given $\bra T_{ij}\ket$ we derive from it, up to zero modes, a transverse, traceless tensor 
$\ti h_{ij}$ of dimension zero:
\be\label{TC1}
\ti h_{ij}={4\ve\over\Box^3}\,C_{ij}(\bra T\ket)~,
\ee
where from now on $\ve=-1$ in Lorentzian signature, and $\ve=1$ in the Euclidean. The two theories are related by
a Legendre transform which is a gravitational version of electric-magnetic duality. Thus 
the bulk is dual to two possible CFT's, depending
on whether the on-shell action is interpreted as the generator of connected correlation functions of the stress-energy
tensor or as the effective action. In the absence of matter, the operators seen by the bulk are the 
correlators of the respective stress-energy tensors. 
 Duality essentially 
interchanges CFT$_1$, with fixed source $h_{ij}$ and stress-energy tensor $\bra T_{ij}\ket$, with CFT$_2$, with 
dual source $\ti h_{ij}$ and stress-energy tensor $\bra\ti T_{ij}\ket$. This dual graviton has opposite parity 
from the original one.

The bulk action connects both CFT's through the coupling \eq{set}. Using \eq{TC}, the variation \eq{set} produces the Cotton 
tensor and the resulting term becomes of the gravitational Chern-Simons type connecting both gravitons. This is the
term that generates the Legendre transformation. This is 
the same result as for abelian gauge fields, where the effective action is a $BF$-term that defines the 
$S$-duality operation. Notice that all this is true independent of the linearization.

The two fixed gravitons are related by the $S$-duality operation that will be defined later,
\be
S(h_{ij})=-\ti h_{ij}~.
\ee
This operation squares to minus one. It performs a Legendre transformation on the action, and hence modifies the
holographic dictionary. The usual dictionary identifies the on-shell action with the generating functional of
connected correlation functions of the stress-energy tensor in CFT$_1$. After $S$-duality, the bulk on-shell 
action can be defined to give the boundary effective action of CFT$_2$ instead.

There are mixed boundary conditions corresponding to coupling the theory to dynamical gravity and integrating over
the graviton. To illustrate this, let us consider a slightly intermediate problem where the boundary theory is only
coupled to Cotton gravity, and one integrates over the conformal metrics plus other degrees of freedom of the
CFT. The mixed boundary condition corresponding to this fixes a linear combination of the boundary metric and the 
stress-energy tensor of the original theory. This means adding to the bulk action a gravitational Chern-Simons term $k/4\pi S_{\sm{CS}}$
(defined in Appendix \ref{appC}) and taking the variation:
\be
\d S=({\mbox{eom}})+\half\int\dd^3x\,\sqrt{g}\left({3\ell^2\over2\k^2}\,g_{(3)ij}-{k\over4\pi}\, C_{ij}(g)\right)\d g^{ij}~,
\ee
and recall that $g_{(3)}$ is the stress-energy tensor of the unperturbed Dirichlet theory, \eq{stressT}.
The mixed boundary condition reads
\be\label{mixedTC}
g_{(3)ij}={k\k^2\over6\pi\ell^2}\,C_{ij}
\ee
where $C_{ij}$ is the Cotton tensor. Linearizing, the analytic continuation of the Euclidean regularity condition
(which will be derived in section 2) $\bar h_{(3)}={1\over3}\Box^{3/2}\bar h_{(0)}$ gives:
\be\label{SD0}
\Box^{1/2}\bar h_{(0)}=-\left(\k\over\ell\right)^2{k\over6\pi}\,\e_{ikl}\pa_k\bar h_{(0)jl}~.
\ee
By dualizing this equation, we find that the coefficient on the right-hand side must be $\pm1$, and hence 
that the coefficient of the Chern-Simons term is given in terms of Newton's constant as:
\be\label{fixed}
k=\pm6\pi\left(\ell\over\k\right)^2~.
\ee
This value also corresponds to the gravitational instantons recently found in \cite{STproc,ST}. We will show that
equation \eq{SD0}
has solutions with non-trivial graviton profile. They are specified by a single parameter, which determines the 
value of the action, and in that sense they have no dynamics. The above illustrates the fact that the boundary
graviton may have non-trivial configurations. 

Obviously, \eq{SD0} also admits $k=0$ and $k=\infty$. $k=0$ is simply the unperturbed theory with zero stress-energy
tensor. At $k=\infty$ the gravitational Chern-Simons term dominates and we get Cotton gravity, i.e. zero 
Cotton tensor. This sets the transverse, traceless part of the graviton fluctuations to zero, thus getting back the
Dirichlet boundary condition for the transverse, traceless graviton. For generic $k$ as in \eq{fixed}, $g_{(3)}$ acts as a stress-energy tensor and the original 
CFT is coupled to Cotton gravity. Although the 
perturbation considered here is classically marginal, we do not get a marginal line of boundary conditions,
 but rather marginal points \eq{fixed}, $k=0$, and $k=\infty$.

Instanton solutions \eq{SD0} are topological, but one can get proper boundary dynamics by considering 
non-marginal perturbations, if one couples the CFT to dynamical gravity. The additional points \eq{fixed} of the
topological situation become a line and we are left with only two ``fixed points'', $k=0$, $k=\infty$.

The degrees of freedom of boundary 
gravitons are then best described by two scalars $\g$ and $\psi$,
in the following way. It turns out that the transverse, traceless part of the graviton can be written as:
\be\label{bdyg}
\bar h_{ij}(r,p)=\g(r,p)\,E_{ij}(p)-i\psi(r,p)\, D_{ij}(p)
\ee
where
\bea\label{defED}
E_{ij}&=&{\bar p_i^*\bar p_j^*\over\bar p^{*2}}-\half\Pi_{ij}\nn
D_{ij}&=&{1\over2p^3}\left(\bar p_i^*\e_{jkl}+\bar p_j^*\e_{ikl}\right)p_k\bar p_l^*
\eea
are the two unique transverse, traceless tensors of rank two that can be constructed from $p,p^*$. They are each
other's duals and will play an important role in the following, as they determine the index structure and duality 
properties of the metric. 

The dual graviton $\ti h$ has the same expansion as $h$:
\be
\ti h_{ij}(r,p)=\ti \g(r,p)\,E_{ij}(p)-i\ti\psi(r,p)\, D_{ij}(p)~,
\ee
and duality interchanges $\g\leftrightarrow\ti\g$, $\psi\leftrightarrow\ti\psi$.

For generic Robin boundary conditions, the coupling between the two gravitons leads to a theory with non-zero
stress energy tensor even in the absence of a source, whereas the vev in the original theory is zero.

The content of the paper is as follows. In section 2 we give the bulk solutions that are used later,
including regularity of the Euclidean solution and the value of the on-shell action. In section 3, mixed 
boundary conditions are discussed and it is shown that they give rise to topologically massive gravity on the
boundary. We solve the instanton equations explicitly and compute the on-shell value of the action. Section 4 is
the main section, where we work out gravitational electric-magnetic duality: we show that it follows from a symmetry
of the equations of motion, show that it acts as a Legendre transformation, we construct the dual graviton at the 
non-linear level, give the bulk interpretation and we discuss holography of the mixed boundary conditions. We
find that the stress-energy tensor can spontaneously acquire a non-zero vev. We also compute two-point
functions and discuss cosmological topologically massive gravity. In section 5 we discuss our results and give some
future directions.

As this work was being completed, a nice preprint \cite{CM} appeared dealing with the first problem mentioned
before. The results 
by Ishibashi and Wald were found to hold for $d\leq4$ and were conjectured to hold for higher dimensions as well, 
the key issue being to take into account the counterterms that render the action finite in the definition of the symplectic structure.
Very recent related work also appeared in \cite{MPT}.

\section{Bulk dynamics}

We will solve Einstein's equations perturbatively about the AdS$_4$ background. We take the standard 
Fefferman-Graham form of the metric:
\bea
\dd s^2&=&{\ell^2\over r^2}\left(\dd r^2+g_{ij}(r,x)\dd x^i\dd x^j\right)\nn
g_{ij}(r,x)&=&\eta_{ij}+h_{ij}(r,x)~,
\eea
where the metric fluctuations have the following expansion:
\be\label{FGexp}
h_{ij}(r,x)=h_{(0)ij}(x)+r^2h_{(2)ij}(x)+r^3h_{(3)ij}+\ldots
\ee
In four dimensions there are no logarithmic terms and one can show that the linear term is absent.
We have fixed the gauge $h_{rr}=h_{ir}=0$. We will raise and lower indices with $\eta_{ij}$ ($\d_{ij}$ in the
Euclidean) and denote $h=h_{ii}$. Einstein's equations about the linearized background take the form:
\bea\label{linear}
h''-{1\over r}\,h'&=&0\nn
\pa_jh'_{ij}-\pa_ih'&=&0\nn
h_{ij}''-{2\over r}\,h_{ij}'+\Box h_{ij}+\pa_i\pa_j-\pa_i\pa_kh_{jk}-\pa_j\pa_kh_{ik}-{1\over r}\,h'\eta_{ij}&=&0~,
\eea
where the primes denote $r$-derivatives. We immediately find that the trace and transverse parts of the graviton
are quadratic in $r$,
\bea
h(r,x)&=&h_{(0)}(x)+r^2\,h_{(2)}(x)\nn
\pa_jh_{ij}(r,x)&=&\pa_jh_{(0)ij}(x)+r^2\,\pa_jh_{(2)ij}(x)~,
\eea
and all higher-order coefficients are conserved and traceless, $h_{(n)ii}=\pa_jh_{(n)ij}=0$ for $n>2$. Using this,
we can rewrite the last line in \eq{linear} as an equation for purely the transverse, traceless part of the graviton 
defined in \eq{tt}:
\be
\bar h_{ij}'-{2\over r}\,\bar h_{ij}'+\Box\bar h_{ij}=0~.
\ee
In Euclidean signature, the general solution is:
\be\label{hEucl}
\bar h_{ij}(r,p)=f_1(r,p)\,a_{ij}(p)+f_{(3)}(r,p)\,b_{ij}(p)~,
\ee
where $f_1$ and $f_3$ are the two independent solutions of the above equation written in momentum space:
\bea
f_1(r,p)&=&\cosh pr-pr \sinh pr\nn
f_3(r,p)&=&-\sinh pr+pr\cosh pr~.
\eea
Near the boundary we recover the expansion \eq{FGexp} in the distance to the boundary:
\bea
\bar h_{ij}(r,p)&=&\left(1-\half\,p^2r^2+\ldots\right)\bar h_{(0)ij}(p)+(r^3+\ldots)\,\bar h_{(3)ij}\nn
\bar h_{(0)ij}(p)&=&a_{ij}(p)\nn
\bar h_{(2)ij}(p)&=&-\half\,p^2\bar h_{(0)ij}\nn
\bar h_{(3)ij}(p)&=&{p^3\over3}\,b_{ij}(p)~.
\eea
The series decomposes into two independent series with even and odd powers of $r$.
$h_{(0)}$ multiplies even powers and $h_{(3)}$ multiplies the odd powers. By \eq{stressT}, $h_{(3)}$ is
the holographic stress-energy tensor.
The remaining non-transverse, traceful parts of the metric are given in \eq{fullh}.

It is convenient for later use to rewrite the above in the form \eq{bdyg} where the two independent components of
the metric are made completely explicit. Demanding that \eq{bdyg} satisfies the bulk equations of motion, i.e. that
it has the form \eq{hEucl} (or its Lorentzian counterpart), we get the following expansion:
\bea
\g(r,p)&=&\g(p)f_0(r,p)+\d(p) f_3(r,p)\nn
\psi(r,p)&=&\psi(p)f_0(r,p)+\chi(p)f_3(r,p)~.
\eea
The first and third coefficients then read:
\bea\label{h0h3}
\bar h_{(0)ij}(p)&=&\g(p)\,E_{ij}(p)-i\psi(p)\,D_{ij}(p)\nn
\bar h_{(3)ij}(p)&=&{\ve\over 3}|p|^3\left(\d(p)\,E_{ij}(p)-i\chi(p)\,D_{ij}(p)\right)~.
\eea
In Appendix \ref{PropED} we give the main properties of the tensors $E,D$.

\subsection{Regularity and on-shell action}

The above Euclidean solution blows up exponentially at $r=\infty$ unless $a_{ij}=b_{ij}$, which gives the regularity condition:
\be\label{reg}
h_{(3)ij}(x)={1\over3}\,|\Box|^{3/2}h_{(0)ij}~,
\ee
hence relating the one-point function of the dual operator to the source, as usual.

In the Lorentzian, the solution oscillates and there is no regularity condition of the type found for Euclidean
solutions:
\be\label{hLor}
h_{ij}(r,p)=a_{ij}(p)\left(\cos(|p|r)+|p|r\sin(|p|r)\right)+b_{ij}(p)(|p|r\cos(|p|r)-\sin(|p|r))~.
\ee

The on-shell action can be easily obtained as follows. From the definition of the stress-energy tensor \eq{set},
using $h_{(3)ij}(p)=f(p)\,h_{(0)ij}(p)$ for {\it arbitrary} $f(p)$ (of which the regular solutions \eq{reg} are a 
special case), and integrating, we get
\be\label{genf}
S_{\sm{on-shell}}={3\ell^2\over8\k^2}\int\dd^3x\,\bar h_{(0)ij}\bar h_{(3)ij}=:W[h_{(0)}]~,
\ee
where gauge-dependent terms do not contribute, as can be checked using the results in Appendix \ref{Lorentzian}.  
This expression is independent of the choice of linear boundary condition, that is the choice of $f(p)$; the fact
that the relation is linear uniquely specifies the on-shell value of the action up to quadratic order \cite{SdHPeng}. 
The same result is obtained expanding the action as in \cite{GlebSergei} and subtracting the divergent part 
with the usual counterterms \cite{dHSS}. 

\section{Dynamical gravity on the boundary}

One important difference between fields of spin 1 and 2, and scalar fields in AdS$_4$, is that the natural linear
boundary conditions preserving conformal invariance involve derivatives rather than being algebraic. 
Generalized boundary conditions involving derivatives therefore may have zero modes that propagate purely on the
boundary and lead to boundary degrees of freedom. This was first pointed out for abelian gauge fields in 
\cite{SdHPeng}; for gravity, it was pointed out in \cite{SdHtalk} and \cite{CM}. Instanton boundary conditions for 
$U(1)$ gauge fields involve a parity-odd operator which is linear in spatial derivatives \cite{SdHPeng}; 
instanton boundary conditions for 
the graviton involve a parity-odd, dimension three operator, the Cotton tensor, and they correspond to coupling 
the boundary theory to Cotton gravity, as explained in the introduction. More general boundary conditions may 
be obtained by coupling the theory to Einstein gravity. This will be worked out in this section and the next.

\subsection{Linear boundary conditions}

The simplest linear boundary condition reads:
\be\label{SD}
\Box^{1/2}\bar h_{(0)ij}=\pm\,\e_{ikl}\pa_k\bar h_{(0)jl}~.
\ee
As in \eq{mixedTC}, this can be obtained adding a gravitational Chern-Simons term to the action. It may also be
obtained from bulk configurations as in \cite{STproc,ST}, by considering solutions with self-dual Weyl tensor:
\be\label{SDW}
C_{\m\n\a\b}=\half\,\e_{\m\n\l\s}\,C_{\a\b}{}^{\l\s}~.
\ee
Asymptotically, this leads to \eq{SD}.

The linear boundary condition \eq{SD} can be written in terms of the curvature tensor
as
\be
\bar R_{ij}=\mp{1\over\Box^{1/2}}\,C_{ij}~,
\ee
where $\bar R_{ij}=\Pi_{ij}{}^{kl}R_{kl}$, and $\Pi_{ijkl}$ is the spin-two projector defined in Appendix 
\ref{curvtensors}. Hence, only the transverse, traceless part of the curvature is involved.
We will now solve this equation. It is useful to first rewrite it as a self-duality
equation
\be\label{SDd}
h_{ij}=\pm d_{ij}[h]
\ee
for some operator $d$. One can formally give the following non-local expression for it:
\be
d={2\over{(-\ve \Box)}^{3/2}}\,C~,
\ee
where $C$ is the Cotton tensor. Notice that the operator in the denominator is positive definite in both signatures.
One easily checks that $d$ squares to one,
\be
d^2=1~.
\ee
Local expressions are obtained by multiplying both sides with the Laplacian; the above definition, though
formal, is very useful for book-keeping. 

The most general self-dual solution of \eq{SD} then takes the following form:
\be\label{SDsol}
\bar h_{ij}=E_{ij}\pm d_{ij}[E]~,
\ee
where $E_{ij}$ is a general transverse, traceless tensor. By construction, the solution is (anti)self-dual. 
$E_{ij}$ can be constructed uniquely up to an overall factor. This is done in Appendix \ref{appB}. In momentum 
space, up to an overall factor it takes the form:
\be
E_{ij}={\bar p_i^*\bar p_j^*\over\bar p^{*2}}-\half\Pi_{ij}~.
\ee

In general, we may expand the graviton as
\be\label{grav}
\bar h_{(0)ij}=\g E_{ij}+{1\over p}\psi'\e_{ikl}p_kE_{jl}
\ee
where $\bar p_i^*$ is the transverse projection of $p_i^*=(-E,\vec p)$, and $\Pi_{ij}$ is the transverse projector.

\subsection{Topologically massive gravity}

Instanton boundary conditions give the traceless, transverse part of three-dimensional topologically massive
gravity, \eq{SD}. We can get the full topologically massive gravity by coupling it to Einstein gravity with 
Newton's constant $1/\m$. The boundary condition for the resulting Neumann problem is:
\be
R_{ij}[g_{(0)}]-{1\over\m}\,C_{ij}[g_{(0)}]={3\ell^2\over4\m\k^2}\,g_{(3)ij}~,
\ee
where we have used the fact that the curvature scalar is zero, $R=0$.
$g_{(3)ij}$ acts here as a source for Einstein's equations. It is as a function of the boundary metric $g_{(0)}$ 
determined by the particular CFT state. At the linearized level, and in the ground state, its functional form is 
given by analytic continuation from the regular Euclidean solutions \eq{reg}. For general states, and at the 
non-linear level, the relation will differ from that. 

In the purely Neumann quantization scheme the stress-tensor vanishes while still getting interesting boundary dynamics. For simplicity we now set $g_{(3)}=0$. It is easy to check that by virtue of $R=0$ the theory is, at the non-linear level, described by a conserved, traceless tensor --the curvature-- satisfying:
\be
\left(\Box+\e\m^2\right)R_{ij}-3\left(R^2_{ij}-{1\over3}\,g_{ij}\Tr(R^2)\right)=0~,
\ee
where $R^2_{ij}:=R_{ik}R^k_j$. Later on this equation will be generalized to the cosmological setting.

\subsection{On-shell action for instantons}

Instantons in gauge theory have two important properties. They are topological, i.e. they have vanishing 
stress-energy tensor and zero Hamiltonian; secondly, their topological class is characterized by an integer $k$ 
giving the on-shell value of the action. 

Bulk gravitational instantons of the type \eq{SDW} have similar properties as we will now show. The self-duality 
equation \eq{SD} imposes $\psi'=\pm\g$ in \eq{grav}, therefore
\be
\bar h_{(0)ij}=\g\left(E_{ij}\pm{1\over p}\,\e_{ikl}p_kE_{jl}\right)~,
\ee
where the metric is (anti-)self-dual.

The on-shell action can be easily computed from the bulk up to second order in the perturbations. Notice that
for a general solution \eq{grav}, we have
\be
h_{(0)ij}(p)h_{(0)ij}(-p)=\half(\g^2+\ve\psi'{}^2)~,
\ee
where $\ve=1$ in the Euclidean, $\ve=-1$ in Lorentzian signature. The Euclidean on-shell action is
\be\label{onshellaction}
S_{\sm{on-shell}}={3\ell^2\over8\k^2}\int\dd^3x\,\bar h_{(0)ij}(x)\bar h_{(3)ij}(x)={\pi^2\ell^2\over8G_N}\,K~,
\ee
with $K=\int\dd^3p\,|p|^3\,\g(p)\g(-p)$. $\g$ should be chosen such that this integral is finite. On the other 
hand, the Hamiltonian is zero. This result is similar to the one for $U(1)$ gauge fields \cite{SdHPeng} in 
AdS$_4$. It would be interesting to see whether there are any topological restrictions on the possible values of
$K$.

\section{Duality}

We have shown that in three dimensions, given a stress-energy tensor $\bra T_{ij}\ket_h$, we can define a 
graviton via the
Cotton tensor: $\bra T_{ij}\ket_h=C_{ij}[\ti h]$. A priori, this graviton does not satisfy any dynamics but is 
given by the response of the one-point function in a particular state to the fixed graviton source $h_{ij}$. 
Given $h_{ij}$, we can also construct a new conserved, traceless, dimension three tensor using 
the Cotton tensor. The question is whether we may interpret the latter as a stress-energy tensor $\bra\ti T_{ij}\ket$
in some dual theory.

From the bulk point of view, the interchange of $h_{ij}$ and $\ti h_{ij}$ corresponds to a change in boundary conditions from Dirichlet to Neumann (and, as we have seen, there are also massive interpolating solutions).
In this section we will first show that the symmetry interchanging $h_{ij}$ and $\ti h_{ij}$ is a duality of the 
electric-magnetic type, for any value of the radial coordinate $r$ and not only at the boundary. The magnetic variable is the Cotton tensor, and 
the electric variable turns out to be the renormalized holographic stress-energy tensor. We will explicitly show 
the symmetry of the bulk equations of motion that leads to this result. Then we will show that CFT$_1$ and 
CFT$_2$ are related by a Legendre transformation which allows us to identify the generating functional of one theory
with the effective action of the other under the familiar dictionary \cite{SdHTP,dHPP}, thus showing 
$C_{ij}[h]=\bra\ti T_{ij}\ket$.

\subsection{Duality symmetry of the equations of motion}

We will identify the duality symmetry of the bulk equations of motion that allows us to define electric and 
magnetic components of the graviton. It will be useful to first recall the case of abelian gauge fields in the bulk of AdS \cite{SdHPeng}. The solution of the bulk equations of motion for a Maxwell field is:
\be\label{Arp}
\bar A_i(r,p)=A_i(p)\cos(|p|r)+{1\over|p|}\,E_i(p)\sin(|p|r)~.
\ee
In Coulomb gauge, the bulk electric and magnetic fields are
\bea\label{BEbulk}
B_i(r,x)&:=&\e_{ijk}\pa_j\bar A_k(r,x)\nn
E_i(r,x)&:=&\pa_r\bar A_i(r,x)~.
\eea
Given an electric-magnetic change of boundary onditions, $B_i'(x)=-E_i(x)$, $E_i'(x)=B_i(x)$, one readily checks
that the bulk fields \eq{BEbulk} also satisfy:
\bea\label{EB1}
B_i'(r,x)&=&-E_i(r,x)\nn
E_i'(r,x)&=&B_i(r,x)~.
\eea
This is a non-trivial property of the equations of motion. $E_i(x)$ and $B_i(x)$ are two independent quantities
on the boundary, namely the one-point function of a holographic one-point function of a global symmetry current, 
and the (curl of) the corresponding source. Their bulk extensions \eq{BEbulk} are related to each other via the first $r$- and boundary spatial derivatives: $E(r,x)=\pa_r\left({1\over\Box^{1/2}}\,*\dd B(r,x)\right)$.
This implies $B(r,x)=\pa_r\left({1\over\Box^{1/2}}\,*\dd B'(r,x)\right)$, which holds by virtue of the equation
of motion, $B(r,x)=B(p)\cos(|p|r)+|p|A'(p)\sin(|p|r)$, where the $r$-derivative exchanges the two oscillating
branches.

In the gravity case, we need to construct bulk electric and magnetic quantities which when interchanged give 
solutions of the equations of motion. It is a priori not entirely obvious which are the correct quantities, as the 
electric and magnetic boundary conditions interchange just the boundary conditions $h_{(0)}$ and $\ti h_{(0)}$. 
And even though these quantities readily extend to bulk fields $\bar h_{ij}(r,p)=f_0(r,p)\,\bar h_{(0)ij}$, 
$\ti h_{ij}(r,p)=f_3(r,p)\,\ti h_{(0)ij}$ satisfying the equations of motion, electric-magnetic variables should 
belong to the {\it same} CFT. 

Based on what he have already said, one may readily expect that electric-magnetic duality exchanges the 
stress-energy tensor at radius $r$ with the Cotton tensor at radius $r$. Indeed, the pair $(\bra T_{ij}\ket,C_{ij}[\ti h])$, 
rather than $(h_{ij},\ti h_{ij})$, belong to the same CFT. How do we check that this is a symmetry 
of the equations of motion, as in \eq{EB1}?

The Cotton tensor has dimension 3. It is therefore natural to compare it with the third time derivative of the graviton (where ``time'' is the $r$-direction). Writing \eq{hLor} as
\be
\bar h_{ij}[a,b]=a_{ij}(p)(\cos(|p|r)+|p|r\sin(|p|r))+b_{ij}(p)(|p|r\cos(|p|r)-\sin(|p|r))~,
\ee
one easily finds by inspection
\be
\bar h'''_{ij}[-b,a]=|p|^3\bar h_{ij}[a,b]-3\,{|p|\over r}\,\bar h'_{ij}[a,b]~.
\ee
This is the basic symmetry that allows to define electric-magnetic duality in the bulk. Expressing $\bar h_{ij}'''$ in
lower derivatives via the quations of motion, $\bar h_{ij}'''={2\over r^2}\,h_{ij}'-|p|^2\bar h_{ij}'-{2|p|^2\over r}\,\bar h_{ij}$,
we get
\bea\label{sym}
-\bar h_{ij}[-b,a]&=&-{1\over|p|^3r^2}\,\bar h_{ij}'[a,b]+{1\over|p|r}\,\bar h_{ij}[a,b]-{1\over|p|}\,\bar h_{ij}'[a,b]\nn
&=:&{1\over|p|^3}\,P_{ij}[a,b]
\eea
$P_{ij}$ becomes, at the boundary, the stress-energy tensor: $P_{ij}(0,p)=-3h_{(3)}$. In fact, we have for any $r$:
\be
\bra T_{ij}(x)\ket_r=-{\ell^2\over2\k^2}\,P_{ij}(r,x)-{\ell^2\over2\k^2}\,|p|^2\bar h_{ij}'(r,x)~,
\ee
and so they differ by a local term that vanishes at the boundary. Thus, up to this finite renormalization,
$P_{ij}(r,x)$ is the renormalized stress-energy tensor at radius $r$. 

However, \eq{sym} is still not the right symmetry, which should interchange the two gravitons, $h_{(0)}$ and 
$\ti h_{(0)}$ instead of $a_{ij}$ and $b_{ij}$. One easily finds
\bea
2C_{ij}(\bar h[-\ti a,a])&=&-|p|^3P_{ij}[a,\ti a]\nn
2C_{ij}(P[-\ti a,a])&=&+|p|^3\bar h_{ij}[a,\ti a]
\eea
This is the form electric-magnetic duality takes in the bulk. It may be obtained as a combination of the operation 
$d=2C/\Box^{3/2}$ defined earlier, and the discrete operation $s(a)=-b$, $s(b)=a$. We will denote it by $S:=sd$.
It acts as expected:
\bea\label{Sh}
S(\bar h_{(0)})&=&-\ti h_{(0)}\nn
S(\ti h_{(0)})&=&+h_{(0)}~.
\eea
We may now define the electric and magnetic variables
\bea
{\cal E}_{ij}(r,x)&=&-{\ell^2\over2\k^2}\,P_{ij}(r,x)\nn
{\cal B}_{ij}(r,x)&=&{\ell^2\over\k^2}\,C_{ij}[\bar h(r,x)]~,
\eea
such that 
\bea\label{EB}
{\cal E}_{ij}(0,x)&=&\bra T_{ij}(x)\ket\nn
{\cal B}_{ij}(0,x)&=&{\ell^2\over\k^2}\,C_{ij}[\bar h_{(0)}]~.
\eea
$S$-duality then acts as
\bea
S({\cal B})&=&-{\cal E}\nn
S({\cal E})&=&+{\cal B}~,
\eea
and obviously $S^2=-1$. Thus, {\it gravitational $S$-duality interchanges the renormalized stress-energy tensor 
with the Cotton tensor at radius $r$}.

There is a second important possibility, which is defining ${\cal B}(r,x)=2C_{ij}[\bar h]$ without the factor of 
$\ell^2/\k^2$. In that case, $h$ and $\ti h$ still satisfy \eq{Sh}, except the coupling $\ell/\k$ now also 
transforms under $S$-duality as $\ell'/\k'=\pm2\k/\l$. This possibility of inverting Newton's constant was 
pointed out in the electromagnetic case \cite{SdHPeng}, and in the gravitational case in \cite{LP2}. Of course, such a 
transformation brings one out of the supergravity regime, and one may only trust the resulting solution if the original one is exact.

\subsection{The gravitational Legendre transformation}

The analysis of Ishibashi and Wald \cite{IW} implies that, as far as boundary conditions are concerned, the 
four-dimensional graviton behaves in a similar way to scalars in the range of masses $-{d^2\over4}<m^2<-{d^2\over4}+1$ ($d=3$). 
The one-parameter line of mixed boundary conditions mimicks the IR flow of boundary CFT's deformed by 
higher-trace operators. The two fixed points are characterized by two CFT's related by a
Legendre transformation.

The gravity picture is quite different because the CFT is deformed by coupling it to gravity. However, electric-magnetic
duality still relates the two ``fixed points'', the Dirichlet and Neumann problems. We will now show that electric-magnetic duality acts on the action 
as a Legendre transformation. For comparison, it is useful to recall here how this works for scalar fields. In the usual CFT with an operator of dimension $\D_+$, $\f_+$ is the fast decaying bulk mode whereas $\f_-$ is the slowly 
decaying mode which is fixed at the boundary. The bulk on-shell action is holographically identified with the generating functional of the boundary CFT as a function of this mode: $S_{\sm{on-shell}}[\f_-]=-W[J]$ with $\f_-=J$. Then, up to contact terms, we have \cite{dHSS}:
\be
\bra{\cal O}_{(\D_+)}\ket_J=-{\d W[J]\over\d J}=-(\D_+-\D_-)\,\f_+~.
\ee
The dual theory is then obtained by first defining the Legendre transformation \cite{yiannis}:
\be
{\cal W}[\f_-,\f_+]=W[\f_-]+\int\dd^3x\,\sqrt{g_{(0)}}\,\f_+(x)\f_-(x)~.
\ee
Extremizing this functional with respect to $\f_-$, ${\d W\over\d\f_-}+\f_+=0$, we get a solution $\f_-=\f_-[\f_+]$.
The dual generating functional is now defined by evaluating ${\cal W}$ at the extremum so it becomes purely a function of $\f_+$,
\be
\ti W[\f_+]={\cal W}[\f_-[\f_+],\f_+]=W[\f_-]|+\int\dd^3x\,\sqrt{g_{(0)}}\,\f_-\f_+|~,
\ee
where the right-hand side is evaluated at the extremum.
Thus, the generating functional of the dual theory is identified with the effective action of the original theory,
but now $\f_+=\ti J$ is fixed. The dual operator is:
\be
\bra\ti{\cal O}_{(\D_-)}\ket_{\ti J}={\d\ti W[\f_+]\over\d\f_+}=\f_-~.
\ee
In \cite{SdHPeng} it was shown that for abelian gauge fields, a Chern-Simons term generates a Legendre transformation
between a theory with a global current of dimension 2, and a theory with a dual current constructed from a dual
gauge field.

The gravitational electric-magnetic analog of the above involves a gravitational Chern-Simons term, as we show next. In the usual CFT, $T_{ij}={2\over\sqrt{g}}{\d W\over\d g^{ij}}$ where $W[g]$ is the on-shell bulk action. We construct the dual CFT in the usual way:
\be\label{calW}
{\cal W}[g,\ti g]=W[g]+V[g,\ti g]~.
\ee
We want to calculate $V[g,\ti g]$. At the extremum, ${\d{\cal W}\over\d g^{ij}}=0$, we have 
${1\over\sqrt{g}}{\d V\over\d g^{ij}}=-\half\bra T_{ij}\ket$, where the last variation is simply a partial derivative.
Linearizing and using the fact that $\bra T_{ij}\ket$ can be written as: 
\be\label{dualizeT}
\bra T_{ij}\ket={\ell^2\over\k^2}\,C_{ij}[\ti h]~,
\ee
we get
\be
V[h,\ti h]=-{\ell^2\over2\k^2}\int\dd^3x\,h^{ij}C_{ij}[\ti h]=-{\ell^2\over2\k^2}\int\dd^3x\,\ti h^{ij}C_{ij}[h]~.
\ee
Since $C_{ij}={1\over\sqrt{g}}{\d S_{\sm{cs}}\over\d g^{ij}}$, $V$ can be rewritten as the variation of the
gravitational Chern-Simons term:
\be\label{CSAB}
V[h,\ti h]=-{\ell^2\over2\k^2}\int\dd^3x\,h^{ij}{\d^2S_{\sm{CS}}[g]\over\d g^{ij}\d g^{kl}}\,\ti h^{kl}=:S_{\sm{CS}}[h,\ti h]~.
\ee

We may now define
\be\label{tiW}
\ti W[\ti h]=W[h]+V[h,\ti h]~,
\ee
where $h$ is an extremum of the action. In the case at hand, using \eq{genf}:
\be
W[h]={\ell^2\over8\k^2}\int\dd^3x\,h_{(0)ij}\Box^{3/2}h_{(0)ij}~,
\ee
we get back the regularity condition $h_{(3)}={1\over3}\,\Box^{3/2}h_{(0)}$, as we should. The latter is rewritten
as $C_{ij}(\ti h)=\half\Box^{3/2}h_{ij}$. In this case, the dual functional is:
\be\label{dualfunctional}
\ti W[\ti h]=-{\ell^2\over8\k^2}\int\dd^3x\,\ti h_{(0)ij}\Box^{3/2}\ti h_{(0)ij}~.
\ee

We can now compute the dual stress-energy tensor:
\be\label{Tdual}
\bra \ti T_{ij}\ket=-2\,{\d\ti W[\ti h]\over\d\ti h^{ij}}={\ell^2\over\k^2}\,C_{ij}[g]~.
\ee

From \eq{EB}, we find that the dual stress-energy tensor is actually the magnetic field:
\be
\bra\ti T_{ij}(x)\ket_{\sm{CFT}_2}={\cal B}_{ij}(0,x)={\cal E}_{ij}'(0,x)=S\left(\bra T_{ij}(x)\ket_{\sm{CFT}_1}\right)~,
\ee
and this definition extends in the natural way to any finite $r$.

The same result can be obtained by applying $S$-duality \eq{Sh} directly on 
the generating functional \eq{genf}:
\be
W'[\ti h]:=S(W[h])=W[h]-{\ell^2\over2\k^2}\int\dd^3x\,\bar h_{ij}C_{ij}[\ti h]~.
\ee

Given that the relation between the generating functionals of CFT$_1$ and CFT$_2$ is a Legendre transformation, 
and since the graviton in one theory becomes after differentiation the stress-energy tensor of the dual theory, 
we may identify the generating functional of one theory with the effective action of the other:
\bea
\G[\bra T_{ij}\ket(\ti h)]&=&+\ti W[\ti h]\nn
\ti \G[\bra \ti T_{ij}\ket(h)]&=&-W[h]~.
\eea
Thus, in the dual theory the holographic dictionary is modified. This modification is as in the case of 
duality for spin-zero and spin-one fields in the special range of masses \cite{SdHTP,SdHPeng}.

\subsection{The non-linear dual graviton}

The above considerations can be extended to the non-linear theory. Let us first consider the issue of non-linear
boundary conditions. Generic non-linear Neumann boundary conditions take the form:
\be\label{NLbc}
{3\ell^2\over4\k^2}\,g_{(3)ij}={\ell^2\over2\k^2}\,C_{ij}[\ti g]=
\m\left(R_{ij}[g]-\half\,g_{ij}R[g]+\l\,g_{ij}\right)-C_{ij}[g]~.
\ee
$\l$ is a boundary cosmological constant, which we will take to be positive $\l>0$. As we will now show, this
equation arises as a Neumann boundary condition of the bulk action deformed by boundary terms; therefore, it
should be read as determining $g_{(3)}$ in terms of $g_{(0)}$. It must be supplemented
by a regularity condition (in Euclidean signature) or by a specification of the state (in Lorentzian signature).
These may take the form of a relation between $g_{(3)}$ and $g_{(0)}$, as we found in the linear case (see e.g.
\eq{reg}); in the non-linear case this relation will be very involved (see \cite{KSBCvR} for progress in this
direction). In turn, because \eq{NLbc} is a differential rather than an algebraic equation, the metric $g_{(0)}$
satisfies a dynamical equation. For $g_{(3)}=0$, this is the topologically massive gravity we encountered 
earlier.
Thus, we modify the action with terms:
\be
S=S_{\sm{EH}}+{\m\ell^2\over4\k^2}\int\dd^3x\sqrt{g}\left(R[g]-2\l\right)-{\ell^2\over4\k^2}\,S_{\sm{cs}}~,
\ee
and $S_{\sm{EH}}$ is the bulk Einstein-Hilbert action, including the Hawking-Gibbons term and counterterms, given in 
\eq{EHaction}.

It is important to notice the role of the bulk Einstein-Hilbert term here. Effectively, it acts as a Chern-Simons
coupling between $\ti g$ and $g$. Indeed, using the non-linear version of \eq{dualizeT}, we find in the on-shell action:
\be
\d S_{\sm{EH}}={\ell^2\over2\k^2}\int\dd^3x\sqrt{g}\,\d g^{ij}C_{ij}[\ti g]~.
\ee
Thus, the bulk produces for us a conformal, diffeomorphism invariant, local Lorentz invariant and fully non-linear
coupling between the two gravitons. This is analogous to 
the situation with (purely linear) Abelian gauge fields, see the discussion in \cite{wittenabelian,SdHPeng}.

For non-abelian theories, it is well-known that there are obstructions to the classical electric-magnetic duality. 
As a prerequisite for the existence of a duality transformation at the non-linear level in the gravity case, it is 
crucial that the existence of the dual graviton $\ti g_{ij}$ can be established beyond the linear 
approximation. The former question will not be addressed in this paper but we will now argue that the latter is 
the case.

Given a dual graviton $\ti g_{ij}$ we can always construct a dual stress-energy tensor using the non-linear 
Cotton tensor:
\be\label{NLC}
\bra T_{ij}\ket={\ell^2\over\k^2}\,C_{ij}[\ti g]~,
\ee
which is automatically traceless and conserved. We need to answer the opposite question:
given the stress tensor $\bra T_{ij}\ket$, can we associate to it a dual graviton $\ti g_{ij}$? In other words, 
is \eq{NLC} true for any CFT? This is the case if the Cotton tensor can be inverted, which up to
zero modes is perturbatively the case around an appropriate background. Dualizing \eq{NLC}, we get
\be
\ve\,\ti\e_i{}^{kl}\ti\nabla_k\bra T_{lj}\ket=-\ti\Box\ti R_{ij}+{1\over4}\ti\na_i\ti \na_j\ti R+{1\over4}\,\ti g_{ij}
\ti\Box\ti R
+3\ti R_{ij}^2-\ti g_{ij}\Tr\ti R^2-{1\over2}\left(3\ti R_{ij}-\ti g_{ij}\ti R\right)\ti R~.
\ee
This equation should be regarded as determining $\ti g_{ij}$, given the left-hand side. Finding $\ti g_{ij}$
is thus equivalent to solving Einstein's equations. We had already established that, at the linear order, the
above equation has solutions, given by \eq{TC1}. The above equation now also gives the higher
order terms in a systematic way. Thus, if perhaps not unique in the presence of zero modes, $\ti g$ exists at the 
non-linear level.

\subsection{Bulk interpretation}

The role of the Chern-Simons term \eq{CSAB}, like the $AB$-Chern Simons terms of \cite{wittenabelian}, is to 
Legendre transform a theory with stress-energy tensor $\bra T_{ij}\ket$ where $h_{ij}$ is fixed, into a
theory with stress-energy tensor $\bra\ti T_{ij}\ket$ where $\ti h_{ij}$ is fixed. From the bulk point of view,
it transforms the Dirichlet problem into a Neumann boundary problem, as we have already seen. It is instructive to
rewrite the above formulas in terms of the bulk quantities:
\be
Z[g]=\int_g{\cal D}G_{\m\n}\, e^{-S[G]}~,
\ee
such that the metric at the boundary approaches $g_{ij}$. Expanding about a flat boundary metric, taking the
Legendre transform and using \eq{calW}-\eq{tiW} gives:
\be
\int{\cal D}h_{ij}\,Z[h]\,e^{V[h,\ti h]}=\int{\cal D}h_{ij}\,e^{{\cal W}[h,\ti h]}=e^{\ti W[\ti h]}=:\ti Z[\ti h]~,
\ee
where we have taken the saddle-point approximation for ${\cal W}[h,\ti h]$, obtaining the dual functional. Thus,
we find
\be
\ti Z[\ti h]=\int{\cal D}h_{ij}\,e^{S_{\sm{cs}}[h,\ti h]}\,Z[h]~,
\ee
and the gravitational Chern-Simons term indeed transforms from Dirichlet to Neumann. This relation is of course
inverted by Legendre transforming back:
\be
\ti{\cal W}[h,\ti h]:=\ti W[\ti h]-V[h,\ti h]=-{\ell^2\over8\k^2}\int\,\ti h^{ij}\Box^{3/2}\ti h_{ij}
+{\ell^2\over2\k^2}\int\,\ti h^{ij}C_{ij}[h]~.
\ee
At the extremum, ${\d\ti{\cal W}\over\d\ti h^{ij}}=0$, we again get the regularity condition, which then gives
$\ti{\cal W}|={\ell^2\over8\k^2}\int\,h^{ij}\Box^{3/2}h_{ij}=W[h]$,
so 
\be
\int{\cal D}\ti h_{ij}\,e^{-S_{\sm{cs}}[h,\ti h]}\,\ti Z[\ti h]=\int{\cal D}\ti h_{ij}\,e^{\ti{\cal W}[h,\ti h]}=e^{W[h]}=Z[h]~,
\ee
as it should.

\subsection{Mixed boundary conditions}

We have shown that, depending on the boundary conditions, we can interpret the same bulk theory either as a 
functional $Z[h]$, where the graviton $h_{ij}$ is fixed and is a source for the stress-energy tensor $\bra T_{ij}\ket$, or in
terms of its Legendre transformed functional $\ti Z[\ti h]$ where the dual graviton $\ti h_{ij}$ is a source for
the dual stress-energy tensor. The latter equals the Cotton tensor of the original theory, \eq{Tdual}.

It was shown by Ishibashi and Wald \cite{IW} that a mixed boundary problem again gives rise to a well-defined
quantization problem in the bulk (see also \cite{CM}). It is therefore natural to seek a generalization of the above where we 
replace the functional $W[h]$ by some new functional ${\cal W}[J]$ that depends on a source that is a linear 
combination of $h$ and $\ti h$:
\be\label{linearbc}
J_{ij}(x)=h_{ij}(x)+\l\,\ti h_{ij}(x)~.
\ee
Since we only want to modify the boundary conditions and not the bulk dynamics, we have to identify $W=S_{\sm{on-shell}}$ as before. However, the form of the holographic stress-energy tensor will change, as we need to vary the action with respect to $J_{ij}$ not $h_{ij}$. This means that we need to modify the definition of $V[h,\ti h]$ as this was obtained from the standard stress-energy tensor.
We define:
\be
{\cal W}[h,J]=S_{\sm{on-shell}}[h]+V[h,J]~.
\ee
It is easy to check that we get \eq{linearbc} provided 
$V[h,J]={\ell^2\over2\k^2\l}\int\left(\half h^{ij}-J^{ij}\right)C_{ij}[h]$:
\bea
{\cal W}[h,J]&=&{\ell^2\over2\k^2}\int\left(\half\,h^{ij}C_{ij}[\ti h]+{1\over\l}\left(\half h^{ij}-J^{ij}\right)
C_{ij}[h]\right)\nn
\d{\cal W}&=&-{\ell^2\over2\k^2\l}\int\,C_{ij}[J-h-\l\,\ti h]\,\d h^{ij}=0~,
\eea
indeed giving back \eq{linearbc}. Since we have not changed the bulk solution but only the boundary conditions, 
$\ti h$ is still determined by regularity or whatever other condition is imposed on a particular solution in Lorentzian signature.
In the case of regular solutions, \eq{linearbc} gives:
\be
J_{ij}=h_{ij}+{\l\over(-\ve\Box)^{3/2}}\,2C_{ij}[h]~.
\ee
This determines $h_{ij}$ in terms of the source, however not completely. There may be zero modes
\be\label{h0ij}
h_{0ij}+{\l\over(-\ve\Box)^{3/2}}\,2C_{ij}[h_0]=h_{0ij}+\l\,d_{ij}[h_0]=0~.
\ee
We recognize here the self-duality equation \eq{SDd}, and we know that its only non-zero solutions are for 
$\l=\pm1$. Thus, we will distinguish the cases $\l=\pm1$ and $\l\not=\pm1$.\footnote{Recently, a similar phenomenon
has been found in the context of AdS$_3$, where $\l=\pm1$ was called the chiral point \cite{LSS}. Here, we find extra degrees of freedom at $\l=\pm1$.}

For any $\l$, the dual stress-energy tensor is
\be
\bra\ti T_{ij}\ket_J=-2{\d S\over\d J^{ij}}={\ell^2\over\l\k^2}\,C_{ij}[h]~.
\ee

{\bf The case $\l=\pm1$ and the non-zero stress-energy tensor vev}

The case $\l=\pm1$ brings us back to the instanton equation \eq{SDd}, which we solved in \eq{SDsol}. In
this special case, $J$ can be easily shown to be self/anti-self dual: $J=h\pm d[h]$ implies $J=\pm d[J]$. In turn,
this implies (anti-) self-dualilty of the dual stress-energy tensor. We can solve for $h(J)$, getting: $h=h_0+\half J$.
Setting the source to zero, we get
\be
\bra \ti T_{ij}\ket_{J=0}=\pm{\ell^2\over\k^2}\,C_{ij}[h_0]=-{\ell^2\over2\k^2}(-\ve\Box)^{3/2}h_{0ij}~.
\ee
The one-point function of the dual stress-energy tensor in CFT$_2$ coupled to gravity does not vanish when the source is set to zero. 
This result is not surprising when there are two coupled gravitons. Although by setting $J_{ij}=0$ we have 
removed the graviton source, 
$h_{0ij}$ (the graviton of the original Dirichlet theory) couples to $J_{ij}$. After setting the source to zero,
the theory has non-vanishing stress and shear. However, notice that by construction the stress-energy tensor is
always conserved. Also, the stress-energy tensor is zero if $h_0$ is a conformally flat fluctuation.
In CFT$_1$, on the other hand, by construction we have  $\bra T_{ij}\ket_h=0$ for any $h$. 


{\bf The case $\l\not=\pm1$}

Now the zero-mode $h_0$ disappears and we solve for $h$ completely in terms of the source:
\be
h={1\over1-\l^2}\left(J-\l\,d[J]\right)~.
\ee
The one-point function:
\be
\bra\ti T_{ij}\ket_J=-{\ell^2\over2\k^2(1-\l^2)}(-\ve\Box)^{3/2}\left(J_{ij}-{1\over\l}\,d_{ij}[J]\right)
\ee
indeed vanishes if $J=0$.

A further generalization of \eq{linearbc} can be achieved by introducing a mass parameter $\m$
and requiring:
\be
J_{ij}=h_{(0)ij}+{3\over\m^3}\,h_{(3)ij}~.
\ee
This is analogous to the usual one-parameter family of boundary conditions that one gets for scalar fields when the CFT is augmented by double-trace deformations. The piece we now add to the action is $V[h,J]={\m^3\over2}\int(\half\,h^2+Jh)$. Using regularity, we rewrite the above as
\be
J_{ij}=\left(1+{(-\ve\Box)^{3/2}\over\m^3}\right)\,h_{ij}~,
\ee
and again it has zero modes These are massive gravitons that satisfy $(-\ve\Box)^{3/2}h_0+\m^3h_0=0$. We
then solve
\be
h=h_0+{\m^3\over(-\ve\Box)^{3/2}+\m^3}\,J~.
\ee
Again, the dual stress-energy tensor acquires a non-zero expectation value, even if $J=0$:
\be
\bra\ti T_{ij}\ket_J=-\m^3h_{0ij}-{\m^6\over(-\ve\Box)^{3/2}+\m^3}\,J_{ij}~,
\ee
whereas $\bra T_{ij}\ket_h=0$, as it should.

\subsection{Duality of two-point functions}

Having computed the on-shell action \eq{onshellaction}, and taking into account the gravitational Chern-Simons 
term, we obtain the two-point function of the stress-energy tensor in the standard way:
\be\label{2pt}
\bra T_{ij}T_{kl}\ket={\ell^2\over\k^2}\,|p|^3\,\Pi_{ijkl}+{ip^2\over\m}\,\e_{imn}p_n\Pi_{jmkl}~,
\ee
which has the standard form including the parity-odd term.

The two-point function of the dual stress-energy tensor is computed by differentiating $\ti W[\ti h]$ with 
respect to the dual graviton instead. For standard duality, where $\k/\ell$ is inert, $\ti W[\ti h]$ was explicitly 
computed in \eq{dualfunctional}. The dual two-point function takes the same form as \eq{2pt}.

However, as pointed out at the end of section 4.1, there is a second possibility where the coupling transforms under $S$-duality as $\ell'/\k'=2\k/\ell$. In that case, the two-point function exhibits the behavior found from field theory in \cite{LP} (without including the Chern-Simons term).

All this is analogous to duality for two-point functions of spin-1 currents studied in \cite{SdHPeng}. In that
case, it was found that full duality involves the gauge coupling as well as the $\th$-angle. For gravity, the
analogous coupling is $\m$. It should be straightforward to generalize the bulk analysis for spin one to the case 
of spin two along the lines of \cite{SdHPeng} and check that under duality Newton's constant and the mass $\m$ mix in the 
way predicted from field theory in \cite{LP}.

\subsection{Cosmological topologically massive gravity}

Finally we show that the theory found in \eq{NLbc} can at the non-linear level be described in terms of a 
massive, conserved and traceless tensor $I_{ij}$. Setting the second graviton to zero, we write:
\be
R_{ij}[g]-\half\,g_{ij}\,R[g]+\l\,g_{ij}=\m\,C_{ij}[g]~.
\ee
It automatically follows that the space has constant scalar curvature, $R=6\l$. Therefore we may define a conserved,
traceless tensor from the curvature as follows:
\be
I_{ij}:=R_{ij}[g]-{1\over3}\,g_{ij}\,R[g]=R_{ij}[g]-2\l\,g_{ij}~.
\ee
Notice that $I_{ij}=0$ gives maximally symmetric solutions. The equation of motion can be rewritten as
\be
I_{ij}={1\over\m}\,\e_i{}^{kl}\na_kI_{jl}~.
\ee
One can check that $I_{ij}$ satisfies 
\be
\left(\Box+\ve\m^2-3\l\right)I_{ij}-3\left(I_{ij}^2-{1\over3}\,g_{ij}I^2\right)=0~,
\ee
where $I^2_{ij}=I_{ik}I^k_j$. This equation describes a non-linear massive conserved and traceless tensor. At the
linearized level, the above equation was also found in \cite{CDWW} (see also \cite{GJJ}). Linearizing, we get 
\be
\d I_{ij}={1\over\m}\,\e_i{}^{kl}\na_k\d I_{lj}~.
\ee
In De Donder gauge:
\be
\d I_{ij}=-\half\,\Box h_{ij}+\l\,h_{ij}-\l\,g_{ij}\,h~,
\ee
where $\nabla^j\d I_{ij}=\d I_i^i=0$ and $(\Box+4\l)h=0$.

All of the above are valid for any value of $\l$. Only for $\l>0$ does the holographic analysis of the previous
sections go through; for $\l<0$, part of the bulk has to be excised, the new boundary having a common two-dimensional
bondary with the usual AdS boundary. We leave the study of holography in this interesting case for the future.

\section{Discussion}

We have shown evidence for the existence of two CFT's with different parity being dual to the same bulk system.
This is a generic fact about gravity, and the dual graviton is always present if the stress-energy tensor is 
non-zero. Therefore, we expect {\it any} gravity theory in AdS$_4$ with conformal boundary conditions to be dual 
to either of two CFT's. This type of duality is already known for conformal matter in a fixed AdS$_4$ background, 
therefore it is reasonable to 
expect it to hold for the coupled gravity-matter system. In that case, the holographic stress-energy
tensor satisfies non-trivial Ward identities \cite{dHSS}, therefore only part of it is given by the Cotton tensor. 
The rest of it will presumably contain all of the dual operators present in the CFT.
It would be interesting to study this in detail. The bulk action gives the non-linear coupling between both 
gravitons in a $BF$-term.

The two gravitons found here have different parity. This is reminiscent of the situation in \cite{Peter} where
$E_{11}$ requires the gravitational degrees of freedom to be described in terms of dual fields. For all the dualities
of this type found in AdS$_4$ so far, there is an embedding in M-theory \cite{SdHPeng,dHPP}. However, there is the
important difference that our dual graviton is not a bulk graviton but lives in three dimensions.

Gravity can become dynamical on the boundary of AdS$_4$ by appropriate choice of boundary conditions. Whereas algebraic
boundary conditions generically fully determine the bulk and boundary values of the fields, Robin-type differential
boundary conditions specify the boundary values only up to zero modes. These zero modes satisfy dynamical 
equations which are holographically identified with the dynamics of the boundary graviton. Depending on the boundary
conditions, the CFT is coupled to Cotton gravity or to topologically massive gravity. 
 The results are similar to the case of deformations for
scalar fields and $U(1)$ gauge fields. The latter is dual to the topologically massive gauge theory \cite{SdHPeng}. 
In the scalar case, deforming the action by irrelevant operators, the theory
flowed towards the IR fixed point. 
In the case of bulk gravity studied in this paper, boundary gravitons are fixed in the Dirichlet and Neumann
problems, which are limiting cases of the mixed boundary conditions. Allowing general mixed boundary conditions
amounts to coupling the CFT to Einstein gravity where the graviton becomes dynamical.

We found examples where the coupling between both gravitons spontaneously generates a non-zero vev for the
stress-energy tensor  of
CFT$_2$ coupled to gravity after the source is removed. This background stress-energy tensor is nevertheless
conserved. In CFT$_1$ coupled to gravity, on the other hand, 
the vev is always zero for the boundary conditions that we have considered. In three dimensions there are no conformal
anomalies. However, there are gravitational anomalies in topological theories. It would be interesting
to see whether this effect is related to the gravitational anomaly of the type discussed in \cite{Jones}. 
Notice that the non-vanishing stress-energy tensor found here is parity-odd and proportional to the Cotton tensor. 
It would be worth studying this mechanism in the light of recent developments in the three-dimensional CFT \cite{BL,ABJM}.

The Legendre transformation relating both theories is electric-magnetic duality, the electric field being the 
{\it renormalized} stress-energy tensor and the magnetic field being the Cotton tensor. This leads to a modification
of the holographic dictionary in the dual theory, where the on-shell bulk action can now be defined to be the
effective action of CFT$_2$. Gravitational
electric-magnetic duality in flat space and in spaces with a cosmological constant has been discussed earlier in 
\cite{Peter,JLR}. Our approach differs from earlier works in that we have given the holographic interpretation
of the duality. For this work it is essential that the boundary quantities are renormalized. Another crucial difference with previous 
work is the existence of a dual graviton on the boundary and a dual stress-energy tensor. The 
existence of this graviton was established at the non-linear level.

Duality in the case of $U(1)$ gauge fields in the bulk has had important applications to condensed matter systems
\cite{HKSS}, as it relates materials with completely different properties. 
$S$-duality has been used to make new predictions for the quantum Hall effect in graphene \cite{BD}. One may
expect duality to play an important role  in the non-abelian case as well (see \cite{HHH} where the bulk 
configuration was conjectured to be dual to a superconductor). In this paper we have found the holographic dictionary for gravitational
electric-magnetic duality, and it is natural to ask which condensed matter systems it relates to each other. In particular, it
is suggested here that the two CFT's have different parity, and it would be interesting to understand the
implications of this for the specific materials.

\section*{Acknowledgements}
\addcontentsline{toc}{section}{Acknowledgements}

We thank Roman Jackiw, Anastasios Petkou and Edward Witten for useful discussions and comments.

\appendix

\section{Holographic renormalization and the Neumann problem}\label{holren}

\subsection{Holographic renormalization: Dirichlet problem}

The renormalized Einstein-Hilbert action is:
\bea\label{EHaction}
S_{\sm{EH}}&=&S_{\sm{bulk}}+S_{\sm{GH}}+S_{\sm{ct}}\nn
&=&-{1\over2\k^2}\int\dd^{d+1}x\sqrt{G}\left(R[G]-2\L\right)-{1\over2\k^2}\int\dd^d x\sqrt{\g}\,2K\nn
&&+{1\over2\k^2}\int\dd^dx\sqrt{\g}\left({2(d-1)\over\ell}+{\ell\over d-2}\,R[\g]\right)
\eea
where $\k^2=8\pi G_N$, $\L=-{d(d-1)\over2\ell^2}$ and $\g$ is the induced metric on the boundary and 
$K=\g^{ij}K_{ij}$. For $d\geq4$, further counterterms appear. 

We choose coordinates
\be
\dd s^2=\dd R^2+e^{2R/\ell}g_{ij}\dd x^i\dd x^j={\ell^2\over r^2}\left(\dd r^2+g_{ij}\dd x^i\dd x^j\right)~,
\ee
with $r/\ell=e^{-R/\ell}$ and
\be
K_{ij}=\half\,\pa_R\g_{ij}={\ell\over r^2}\,g_{ij}-{\ell\over2r}\,g_{ij}'~.
\ee

The variation of the action is:
\be
\d(S_{\sm{bulk}}+S_{\sm{GH}})={1\over2\k^2}\int\dd^dx\sqrt{\g}\left(K^{ij}-\g^{ij}K\right)\d\g_{ij}~.
\ee
The remaining variations are
\be
\d S_{\sm{ct}}=-{1\over2\k^2}\int\dd^dx\sqrt{\g}\left({d-1\over\ell}\,\g_{ij}-{\ell\over d-2}\left(R_{ij}[\g]
-\half\,\g_{ij}R[\g]\right)\right)\d\g^{ij}~.
\ee

We can now compute the boundary stress-energy tensor:
\be
\bra T_{ij}\ket={2\over\sqrt{g_{(0)}}}\,{\d S_{\sm{EH}}\over\d g^{ij}_{(0)}}=\lim_{\e\rightarrow0}\left({\ell\over\e}\right)^{d-2}
{2\over\sqrt{\g(\e,x)}}\,{\d S_{\sm{EH}}\over\g^{ij}(\e,x)}
\ee
evaluated at $r=\e$ and taking the limit. We get:
\be
\bra T_{ij}\ket=-{1\over\k^2}\left(\ell\over\e\right)^{d-2}\left[K_{ij}-\g_{ij}K+{d-1\over\ell}\,\g_{ij}
-{\ell\over d-2}\left(R_{ij}[\g]-\half\,\g_{ij}R[\g]\right)\right]
\ee
In three dimensions, this gives
\be\label{stress}
\bra T_{ij}\ket={3\over2}\left(\ell\over\k\right)^2g_{(3)ij}~,
\ee
as it should.

\section{Boundary graviton and dual graviton}\label{appB}

In three dimensions, we may expand the graviton in terms of $p_i$, $p_i^*$ and a third independent polarization
vector $w_i$, as follows:
\be
h_{ij}(r,p)=w_iw_j+b\,(p_i^*w_j+p_j^*w_i)+c\,(p_iw_j+p_jw_i)+d\,(p_i^*p_j+p_j^*p_i)+e\,p_ip_j+\f\,
{p_i^*p_j^*\over\bar p^{*2}}~.
\ee
There is a lot of redundancy in this expression. Obviously, $c,d$, and $e$ are gauge parameters. Furthermore,
the $b$-dependent term can be removed by a shift of $w_i$. However, this is only useful when a single graviton
is around; if there are two gravitons around, the $b$-term cannot be removed. 

We are interested in the transverse, traceless part of $h_{ij}$:
\be\label{hbar}
\bar h_{ij}(r,p)=\Pi_{ijkl}\left(w_kw_l+\f\,{p_i^*p_j^*\over\bar p^{*2}}+bp_k^*w_l+bp_l^*w_k\right)~,
\ee
where we have reabsorbed $a(r,p)$ in $w_i(r,p)$ and rescaled $b(r,p)$ by the same factor.

$w_i(r,p)$ is a two-dimensional vector with fixed direction and arbitrary scale factor. Its transverse part is
$\bar w_i=\Pi_{ij}w_j$. A convenient choice for $w$ is one where $(p_i,\bar p_i^*,\bar w_i)$ form an orthogonal basis:
\bea
\bar w_i(r,p)&=&\sqrt{\vf(r,p)}\,v_i(p)\nn
v_i(p)&=&v\,\e_{ijk}p_j\bar p_k^*~,
\eea
and $v=1/\sqrt{\ve p^2\bar p^{*2}}$ such that $v_iv^i=1$. Defining $\g=\f-\vf$ and $-i\psi=2\sqrt{\vf}bvp^3$ and
filling it in \eq{hbar}, we get the form \eq{bdyg}.

Regular bulk solutions are obtained imposing \eq{reg}. Filling this in \eq{hEucl}, we get:
\bea
\bar h_{ij}(r,p)&=&e^{-|p|r}(1+|p|r)\,\bar h_{ij}(p)\nn
\bar w_i(r,p)&=&e^{-|p|r/2}\sqrt{1+|p|r}\,\bar w_i(p)\nn
\f(r,p)&=&e^{-|p|r}(1+|p|r)\,\f(p)~.
\eea

As we reviewed in \eq{TC}-\eq{TC1}, in three dimensions, given a conserved, traceless stress-energy tensor, it is
always possible to introduce a dual graviton $\ti h_{(0)}$:
\be
h_{(3)ij}(p)=-ip^2\e_{ikl}p_k\ti h_{(0)jl}(p)~,
\ee
again a traceless transverse, dimension zero tensor. $\ti h_{(0)ij}$ extends to a bulk graviton $\ti h_{ij}(r,p)$,
defined as the Cotton tensor of the holographic stress-energy tensor evaluated at cutoff $r$, $\bra T_{ij}\ket_r$.
$\ti h_{(0)}$ has the same expansion as $h_{(0)}$:
\be
\ti h_{(0)ij}=\ti \g(p)E_{ij}(p)-i\ti\psi(p) D_{ij}(p)~.
\ee
Comparing with \eq{h0h3}, we get 
\bea
\ti\psi(p)&=&-{i\over3}{p^2\over\bar p^{*2}}\,\d(p)\nn
\ti\g(p)&=&-{i\over3}{\bar p^{*2}\over p^2}\,\chi(p)
\eea
in Lorentzian signature. In the Euclidean, the factors of $-i$ are absent. Again, these expressions extend to the
bulk in the obvious way.

\subsection{Properties of the tensors $E_{ij}$ and $D_{ij}$}\label{PropED}

The transverse, traceless tensors $E_{ij}$ and $D_{ij}$ are constructed from the independent vectors
$(p_i,\bar p_i^*,\e_{ijk}p_j\bar p_k^*)$ and are defined in \eq{defED}. They are each others duals:
\bea
D_{ij}&=&{\bar p^{*2}\over p^3}\,\e_{ikl}p_kE_{jl}\nn
E_{ij}&=&-{\ve p\over\bar p^{*2}}\,\e_{ikl}p_kD_{jl}~.
\eea
They also satisfy:
\bea
E_{ik}E^k_j&=&{1\over4}\Pi_{ij}\nn
D_{ik}D^k_j&=&-{\ve\over4}\left(\bar p^{*2}\over p^2\right)\Pi_{ij}\nn
E_{ik}D^k_j&=&{1\over4p^2}\left(\bar p_i^*\e_{jkl}-\bar p_j^*\e_{ikl}\right)p_k\bar p_l^*={1\over4}{\bar p^{*2}\over p^2}\,\e_{ijk}p_k\nn
E_{ij}E^{ij}&=&\half\nn
E_{ij}D^{ij}&=&0\nn
D_{ij}D^{ij}&=&{\ve\over2}\left(\bar p^{*2}\over p^2\right)^2~.
\eea

\section{Three-dimensional curvature tensors}

\subsection{The Cotton tensor and the gravitational Chern-Simons term}\label{appC}

The derivation of the Cotton tensor from the variation of the gravitational Chern-Simons action is standard. 
Up to a total derivative, the latter can be written either in terms of the spin connection or in terms of the
connection 1-form. We will use the spin connection formulation, see for example \cite{MB,DJT}. Writing:
\be
S_{\sm{cs}}=-{1\over4}\int\Tr\left(\o\wedge\dd\o+{2\over3}\,\o\wedge\o\wedge\o\right)~,
\ee
we get
\be
\d S_{\sm{cs}}=-{1\over2}\int\Tr\left(\d\o\wedge R\right)=-\half\int\e^{ijk}R_{ijl}{}^m\d\G^l_{km}=-\int C^{ij}\d g_{ij}~.
\ee
The last line defines the Cotton tensor,
\be
C_{ij}=\half\e_i{}^{kl}\na_k\left(R_{jl}-{1\over4}\,g_{jl}R\right)~.
\ee

\subsection{Linearized tensors}\label{curvtensors}

The projector is defined as follows. The transverse part of $h_{ij}$ is
\be
h_{ij}^\perp=h_{ij}-{1\over\Box}\left(\pa_i\pa_kh_{jk}+\pa_j\pa_kh_{ik}\right)+{\pa_i\pa_j\over\Box^2}\pa_k\pa_lh_{kl}~.
\ee
The transverse, traceless part is now
\bea\label{tt}
\bar h_{ij}&=&h_{ij}^\perp+\half\left({\pa_i\pa_j\over\Box}-\d_{ij}\right)h^\perp\nn
&=&h_{ij}-{1\over\Box}\left(\pa_i\pa_kh_{jk}+\pa_j\pa_kh_{ik}\right)+\half\,\d_{ij}{\pa_k\pa_l\over\Box}h_{kl}
+\half\,{\pa_i\pa_j\pa_k\pa_lh_{kl}\over\Box^2}+\half\left({\pa_i\pa_j\over\Box}-\d_{ij}\right)h~.
\eea

Explicitly, the spin-2 projector is given in terms of the spin-1 projectors:
\bea
\Pi_{ijkl}&=&\half\left(\Pi_{ik}\Pi_{jl}+\Pi_{il}\Pi_{jk}-\Pi_{ij}\Pi_{kl}\right)\nn
\Pi_{ij}&=&\d_{ij}-{\pa_i\pa_j\over\Box}~.
\eea
Of course, it satisfies
\be
\Pi_{ijmn}\Pi_{mnkl}=\Pi_{ijkl}~.
\ee

The three-dimensional Ricci, Schouten, and Cotton tensors are then
\bea\label{variations}
\d R_{ij}&=&-\half\Box\bar h_{ij}+{1\over4\Box}\left(\pa_i\pa_j+\d_{ij}\Box\right)(\Box h-\pa_k\pa_lh_{kl})\nn
\d R&=&\Box h-\pa_k\pa_lh_{kl}\nn
\d P_{ij}&=&-\half\Box\bar h_{ij}+{1\over4}\pa_i\pa_j\left(h-{\pa_k\pa_k\over\Box}h_{kl}\right)\nn
\d C_{ij}&=&\half\,\e_{ikl}\pa_k\Box\bar h_{jl}~.
\eea
The Cotton tensor depends only on the transverse, traceless part of the graviton.

\section{The Lorentzian solution}\label{Lorentzian}

We will solve the three linearized bulk equations of motion \eq{linear} including the gauge-dependent terms. We expand
\be
h_{ij}(r,x)=\bar h_{ij}(r,x)+\d_{ij}\,\f(r,x)+\pa_i\xi_j(r,x)+\pa_j\xi_i(r,x)~.
\ee
The physical part $\bar h_{ij}(r,x)$ has already been solved for. The trace $\f$ and gauge parameters $\xi_i$ satisfy:
\bea
\pa_k\left(\xi_i''-{1\over r}\,\xi_i'\right)+{3\over2}\,\left(\phi''-{1\over r}\,\f'\right)&=&0\label{linear1}\\
\Box\xi_i'-\pa_i\pa_k\xi_k'-2\pa_i\f'&=&0\label{linear2}\\
\pa_i\xi_j''+\pa_j\xi_i''-{2\over r}\left(\pa_i\xi_j'+\pa_j\xi_i'+\d_{ij}\pa_k\xi_k'\right)+\pa_i\pa_j\f+
\d_{ij}\left(\f''-{5\over r}\,\f'+\Box\f\right)&=&0~.\label{linear3}
\eea

The derivative of \eq{linear2} automatically gives:
\be
\Box\f'=0~.
\ee
Taking linear combinations of \eq{linear2} and filling in \eq{linear1}, we get
\be
\Box P_i=\half\,\pa_i\F
\ee
where $P_i=\xi_i''-{1\over r}\,\xi_i'$, $\F=\f''-{1\over r}\,\f'$. Taking further derivatives, and after some
manipulation, we get
\bea
\Box P_i&=&0\nn
\F&=&\F(r)\nn
\pa_iP_i&=&-{3\over2}\F(r)~.
\eea
In these variables, \eq{linear1} is rewritten as
\be
\pa_iP_i=-{3\over2}\,\F~.
\ee
Using $\F=\F(r)$, taking a further derivative we get
\be
\pa_i\pa_k\xi_k''={1\over r}\,\pa_i\pa_k\xi_k'~.
\ee

We now turn to \eq{linear3}. Taking its trace, we get
\be\label{trace}
{2\over3}\pa_k\left(\xi_k''-{5\over r}\,\xi_k'\right)+\f''-{5\over r}\,\f'+{4\over3}\,\Box\f=0~.
\ee
Filling it back in, we get the traceless equation
\be
\pa_i\xi_j''+\pa_j\xi_i''-{2\over3}\,\d_{ij}\pa_k\xi_k''-{2\over r}\left(\pa_i\xi_j'+\pa_j\xi_i'-{2\over3}\,\d_{ij}\pa_k\xi_k'\right)
+\pa_i\pa_j\f-{1\over3}\,\d_{ij}\Box\f=0~.
\ee
Hitting this equation with $\pa_j$ we get
\be
\Box\xi_i''+{1\over3}\,\pa_i\pa_k\xi_k''-{2\over r}\left(\Box\xi_i'+{1\over3}\,\pa_i\pa_k\xi_k'\right)+{2\over3}\,\pa_i\Box\f=0~.
\ee

Equation \eq{trace} gives
\bea
\pa_k\xi_k'&=&-{3\over2}\,\f'+\half\,r\Box\f\nn
\pa_k\xi_k''&=&-{3\over2}\,\f''+\half\,\Box\f
\eea

Equation \eq{linear2} now reads
\bea
\Box\xi_i'&=&\half\,\pa_i\f'+\half\,r\pa_i\Box\f\nn
\Box\xi_i''&=&\half\,\pa_i\f''+\half\,\pa_i\Box\f
\eea

In order to solve \eq{linear3}, we decompose $\xi_i$. In order to facilitate this, we define
\bea
Q_i&=&\xi_i''-{2\over r}\,\xi_i'\nn
Q&=&\f''-{2\over r}\,\f'
\eea
and rewrite it as
\be
\pa_iQ_j+\pa_jQ_i+\d_{ij}Q+\pa_i\pa_j\f=0~.
\ee
We decompose
\be
Q_i=p_ia+\bar p_i^*b+\bar q_i
\ee
where $\bar q\cdot p=\bar q\cdot p^*=0$. We get:
\bea
a&=&-{i\over2}\,\f\nn
b&=&0\nn
p^2&=&0
\eea
and either $\bar q\sim p$ or $\bar q\sim\bar q$. The choice is immaterial. We will choose $\bar q\sim p$, in that
case $\bar q_i$ in the expansion of $Q_i$ can be included in $a$. So we get from the above the solution
\bea
\f''-{2\over r}\,f'&=&0\nn
\xi_i''-{2\over r}\,\xi_i'&=&-\half\,\pa_i\f\nn
\Box\f=\Box\xi_i&=&0~.
\eea
The solution is as follows:
\bea
\f(r,x)&=&\f_{(0)}(x)+r^3\f_{(3)}(x)\nn
\xi_i(r,x)&=&\xi_{(0)i}(x)+r^2\xi_{(2)i}(x)+r^3\xi_{(3)i}(x)~,
\eea
where
\bea
\xi_{(2)i}&=&{1\over 4}\,\pa_i\f_{(0)}(x)\nn
\pa_k\xi_{(3)k}&=&-{3\over2}\,\f_{(3)}\nn
\Box\f_{(0)}&=&0\nn
\pa_i\f_{(3)}&=&0~.
\eea
Hence, the full expansions of the coefficients \eq{FGexp} are:
\bea\label{fullh}
h_{(0)ij}(x)&=&\bar h_{(0)ij}(x)+\d_{ij}\,\f_{(0)}+\pa_i\xi_{(0)j}+\pa_j\xi_{(0)i}\nn
h_{(2)ij}(x)&=&\bar h_{(2)ij}(x)+\half\,\pa_i\pa_j\f_{(0)}\nn
h_{(3)ij}(x)&=&\bar h_{(3)ij}(x)+\d_{ij}\,\f_{(3)}+\pa_i\xi_{(3)j}+\pa_j\xi_{(3)i}~,
\eea
and of course they satisfy
\bea
h_{(2)ii}=\pa^jh_{(2)ij}&=&0\nn
h_{(3)ii}=\pa^jh_{(3)ij}&=&0~.
\eea

\end{document}